# Optical and Electronic Properties of CdTe Quantum Dots in their Freezed Solid Matrix Phase and Solution Phase

Pooja, Papia Chowdhury[a*]

*Department of Physics and Materials Science and Engineering, Jaypee Institute of Information Technology, Noida 201309, Uttar Pradesh, India*

**Abstract**

The present work deals with the comparison of sizes, optical and electronic properties of COOH functionalized CdTe quantum dots (QDs) in freezed solid polymeric (polyvinyl alcohol (PVA) matrix and in solution phase (water). PVA has been chosen as host material for guest CdTe QDs because of its unique properties like hydrophilicity, good thermo stability, and easy process ability. Experimental absorption, emission, X-Ray diffraction spectra and electronic band gap have been studied by UV-Vis absorption, luminescence and X-Ray diffraction spectroscopy. The smaller size of CdTe QDs in solid PVA polymer matrix (~ 6 nm) and larger band gap of ~ 9.5 eV validates their quantum confinement regime in freezed solid phase. The smaller particle size in solid phase compared to that of the particle size in its solution phase (8 nm) validates the non existence of agglomeration in solid phase. Appearance of high intense and wide luminescence emission in solid form proves the strong candidature of CdTe QDs as promising sensors for today's optoelectronic and biomedical industry.

*Keywords: Quantum Dot, CdTe/PVA thin film,UV-Visbile, Photoluminsence spectroscopy, quantum confinement;*

**1. Introduction**

In recent years, among all types of nanomaterials, semiconductor Quantum dots (QDs) show a great attraction towards research areas due to their size-dependent electronic and optical properties [1,2]. Among all types of available QDs, CdTe QDs have a number of significant properties like high quantum yield (QY), photochemical stability and also ability to be tuned in the visible spectral range (380-740 nm) because of this it is widely used in many fields as: detection of hazardous metal ions [3], light emitting diodes, solar cell [4] and fluorescent biological markers [5]. Normally, colloidal semiconductor QDs have large surface energy so tends to agglomerate in solution [6]. To overcome this problem different stabilizer has been used as: capping agents, surfactants and polymers. These stabilizers normally modify the surface functionalities of QDs and help them to maintain their stable size in their nano order. Among different available stabilizer systems, polymers generally work as a host system to many industrially available QDs. Due to their wide unique physicochemical properties and potential applications, polymer and its composites have attracted a great deal of interest in out of many available stabilizers [7]. The presence of QDs in the solid polymer matrix prevents the QDs from agglomeration and makes the QD storage easier and long as compared to the QDs in liquid medium. The optical outputs of semiconductor QDs in different polymer matrix have a versatile importance in every field of research such as: biosensors [8], energy storage materials [9] and nonlinear optics [10]. When freezed in polymer matrix, QDs provide better optical, thermal, electrical, mechanical properties [11] than the QDs in liquid phase. Also the concentration and size of the QDs in the host polymer matrix can be easily controlled due to their freezed configuration which modify the applicability of QD nano composites for the above mentioned applications [12].

* Corresponding author. fax: +91 120 2400986.
*E-mail address:* papia.chowdhury@jiit.ac.in

The aim of the present work is to prepare COOH functionalized CdTe/PVA thin polymer matrix and to determine the size of CdTe QDs in freezed state within PVA matrix. The aim of the study is also to compare the optical and electronic properties of functionalized CdTe QDs in its solid freezed phase with that of its solution phase. The results obtained from the study will definitely help us to utilize the functionalized CdTe QDs for the manufacture of specific sensing devices in the future medical and optoelectronic industries.

**2. Experimental**

*2.1 Material*

COOH functionalized CdTe QDs, Polyvinyl alcohol (PVA) were purchased from Sigma Aldrich. Deionized water (Millipore) was used for solution preparation.

*2.2 Apparatus*

X-ray diffraction (XRD) patterns were recorded by Bruker D8 Advance diffractometer with CuKα radiation source. UV–Visible absorption spectra were recorded at 300 K by a Perkin Elmer spectrophotometer (model Lambda-35) with a varying slit width in the range 190–1100 nm. The thickness of functionalized CdTe/PVA films by were measured by length Measuring Machine, model no OPAL-1000, Make OKM, Jena with a least count of 0.1 μm having range 0-1000mm. All luminescence measurements were made with a Perkin Elmer spectrophotometer (Model Fluorescence-55) with a varying slit width (excitation slit = 10.0 nm and emission slit = 5 nm) ranging from 200 to 800 nm. The Model LS 55 Series uses a pulsed Xenon lamp as a source of excitation. Deionized water (Millipore) was used for measuring absorption and emission spectra of functionalized CdTe QDs in PVA polymer matrix.

*2.3 Preparation of polymer film*

For the preparation of thin film with QDs, we have followed the simple procedure laid down by B. Suo. et.al. [12]. According to this, 2 gm of PVA is dissolved in 30 ml de-ionized water. After that the prepared solution was then kept at magnetic stirrer for 5 hours at temperature 90 °C. After that the thick PVA solution is formed and this thick PVA solution is casted in a glass petri dish. Finally, to evaporate the water from a thick PVA solution, the solution is kept in oven for heating at 70°C for the time period of 10 hours. Functionalized CdTe/PVA thin film was prepared by mixing 1 micro liter of QDs in prepared PVA solution. The concentration of QDs is $10^{-3}$ M. The thickness of the prepared functionalized CdTe/PVA film was found to be ~210 μm.

3. Results and discussion

*3.1 XRD Analysis*

XRD is used to determine the structure and size of the functionalized CdTe QDs within PVA polymer matrix format. The XRD patterns of Pure PVA and with COOH functionalized CdTe QDs are shown in Fig.1 (a, b) respectively. The first diffraction peak of the pure PVA film was at $2\theta = 19.8°$, which shows its semi crystalline nature [13]. The two another diffraction peaks were at $2\theta = 22.7°$ and $40°$ in the functionalized CdTe/PVA thin film, which can be readily assigned due to the (1 1 1) and (2 2 0) planes for CdTe respectively [14]. To calculate the particle size of CdTe QDs, the Debye-Scherrer [15] formula was used (equation 1).

$$D = K\lambda/\beta \cos\theta \quad \ldots\ldots\ldots\ldots\ldots\ldots\ldots (1)$$

Where D is particle size in nm, θ is the Bragg angle and λ is an X-Ray wavelength. Here β is the FWHM of the given peak, K is Scherrer's constant which is 0.9. The estimated size of CdTe QDs in PVA polymer matrix was about 5.81nm [16] with most probable error is 0.04nm.

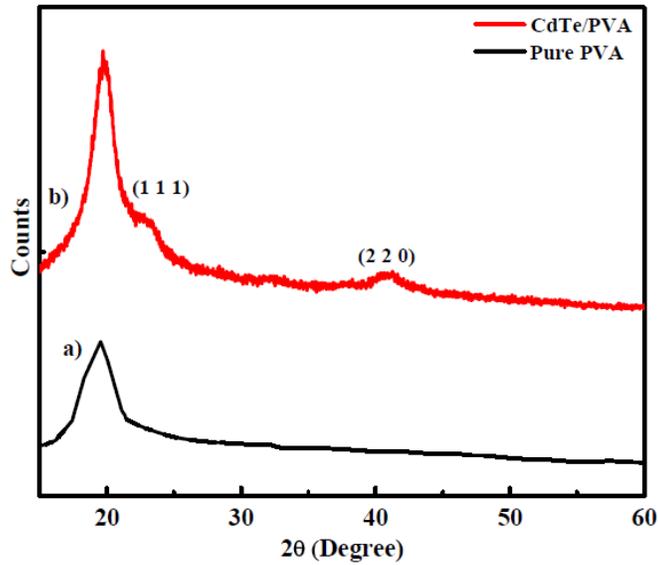

Fig.1 XRD spectra of a) Pure PVA b) CdTe QDs in PVA thin film.

*3.2 Optical properties of CdTe QDs in PVA polymer matrix*

To monitor the particle size and optical properties of functionalized CdTe QDs within PVA polymer mtrix, UV-Vis absorption and photoluminesence data have been studied. The experimental absorption spectrum of PVA thin film and functionalized CdTe QDs in PVA thin film is shown in Fig.2 (a,b). For PVA, the absorption spectrum shows a strong and intense absorption peak at ~ 282 nm which is due to the presence of carboxylic group ($\geq$C=O). After the formation of functionalized CdTe/PVA thin film, a new weak and broad absorption peak is observed at ~ 490 nm (Fig. 3b) with an already existing peak at 282 nm (Fig.2b). The peak of CdTe at 490 nm in presence of polymer matrix is same as it observed for CdTe in solution phase (inset Fig.2.) [17]. So we may comment from the absorption data that PVA is providing an inert background for the CdTe QDs.

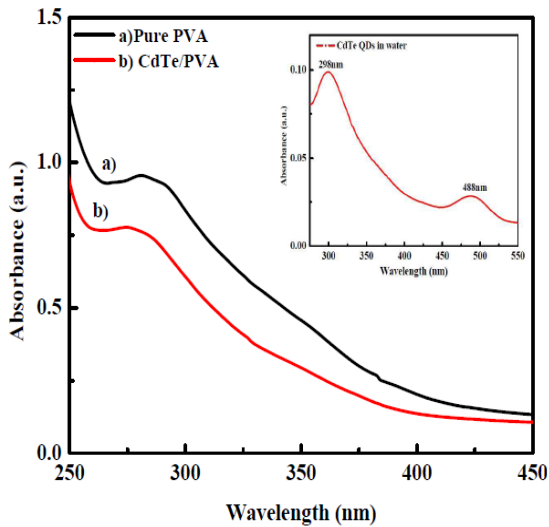

Fig.2. Absorption spectra of a) Pure PVA, b) CdTe QDs in PVA thin film and inset absorption spectrum of CdTe QDs in solution phase.

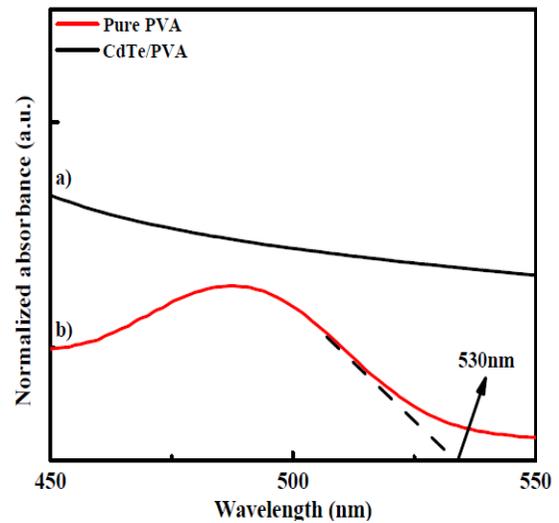

Fig.3. Absorption spectra of a) Pure PVA, b) CdTe QDs in PVA thin film in the wavelength range (450-550 nm).

The average particle size (2R) of prepared functionalized CdTe QDs in PVA thin film can be determined by using the absorption edge of the spectrum by the equation 2 as [18]

$$2R_{(CdTe)} = \frac{0.1}{(0.138 - 0.0002345\lambda_c)} \, nm \ldots \ldots \ldots \ldots (2)$$

where $\lambda_C$ is the absorption edge and 2R is the diameter of the CdTe QD. The absorption edge ($\lambda_C$) can be easily determined by the intersection of sharply decreasing region of the spectrum with the baseline. From this, the estimated size of functionalized CdTe was calculated as ~6.2 nm. In solution phase the estimated size of functionalized CdTe was ~8.3 nm [17]. The close correlation in absorption data of CdTe QDs in liquid phase [17] and functionalized CdTe/PVA in solid matrix phase establishes the signature of existence of CdTe QDs in their solid polymer matrix (Fig.2) also. The observed increment in particle size of QDs in their liquid phase than the solid phase may be due to the agglomeration formation of QDs in liquid phase.

The fluorescence spectra of pure PVA thin film and CdTe in PVA thin film with excitation wavelength at 380 nm are shown in Fig. 4a,b. In case of functionalized CdTe/PVA thin film, we have observed a broad and high intense luminescence band in the range of 520 nm - 540 nm (Fig. 4a) which is same as in the case of CdTe in water medium [17] (inset Fig.4). The Pure PVA shows no emission in this range (Fig. 4a). No emission peak of PVA validates the unperturbed CdTe emission in its solid freezed state in the PVA polymer matrix.

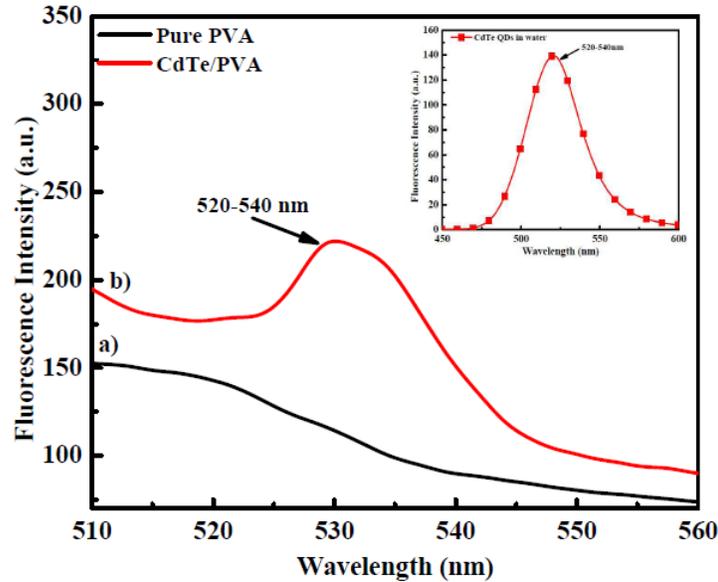

Fig. 4. Photoluminescence spectra of a) Pure PVA, b) CdTe QDs in PVA thin film and inset Photoluminescence spectrum of CdTe in solution phase.

The average particle size (2R) of functionalized CdTe QDs in PVA thin film has been calculated from luminescence data using the following formula (3) [19],

$$D = (9.8127 \times 10^{-7})\lambda^3 - (1.7147 \times 10^{-3})\lambda^2 + (1.0064)\lambda - 194.84 \ldots \ldots \ldots (3)$$

Where, D is the diameter of the QDs in nm and $\lambda$ is the fluorescent emission wavelength of functionalized CdTe/PVA thin film. The particle size of the prepared CdTe QDs in PVA polymer matrix has been estimated as ~ 6 nm which perfectly matches with particle size of QD obtained from absorption data and XRD data (Table 1).

| CdTe QDs | Size | | | Band gap $(\Delta\varepsilon)$ |
|---|---|---|---|---|
| | XRD data | UV-Visible data | Photoluminesence data | |
| Solution Phase | - | 8.3 nm | 8.0nm | 8.63eV |
| Solid Phase | 5.81nm | 6.2 nm | 6.0nm | 9.5eV |

Table1. Size and Band gap of functionalized CdTe/PVA in solution and solid phase.

*3.3 Electronic properties of CdTe QDs in PVA polymer matrix*

The electronic properties of the material can be easily defined by its energy band gap ($\Delta\varepsilon$). Band gap has a great importance in photovoltaic related application [20] and optoelectronic device [21].

$\Delta\varepsilon$ can be calculated from electronic absorption data with the help of famous tauc relation (4) as [22],

$$\alpha h\nu = C(h\nu - E_g)^n \ldots\ldots\ldots\ldots\ldots (4)$$

Where, C is constant, α is an absorption coefficient which can calculate from Beer-Lambert's law, $E_g$ is estimate energy band gap of functionalized CdTe QDs in PVA thin film, n is constant. The values of n = 1/2 and n = 1 for direct and indirect allowed band gap transitions respectively. The average $\Delta\varepsilon$ can be extracted from the intercept of the linear portion of $(\alpha h\nu)^2$ versus $h\nu$ plot (Fig. 5).

$\Delta\varepsilon$ value has been determined in the present work from the experimental absorption data of the prepared functionalized CdTe QD in PVA polymer matrix (Fig. 5). The experimental $\Delta\varepsilon$ is obtained as 9.5 eV (Table 1). The $\Delta\varepsilon$ was larger than the bulk band gap (1.5 eV) of CdTe [23], which must be due to the nano size of the CdTe QDs. The high values of band gap energy also confirm the quantum confinement regime. The $\Delta\varepsilon$ is obtained for CdTe QDs in its liquid phase as 8.63 eV (inset Fig.5) [17] which is smaller than the $\Delta\varepsilon$ observed for CdTe QDs in PVA Polymer matrix. The larger value of $\Delta\varepsilon$ must be due to the decrease in particle size of CdTe QDs in PVA Polymer matrix due to quantum confinement, which is further verified by the XRD, UV-VIS absorption and photoluminescence data.

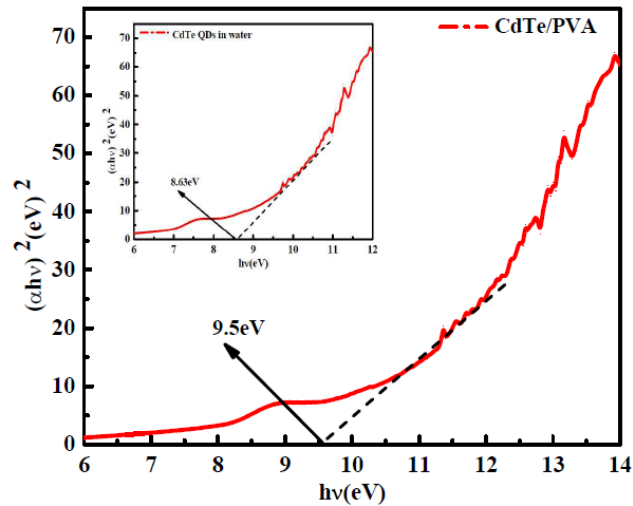

Fig.5. Tauc plot for CdTe QDs in PVA thin film and inset Tauc plot for CdTe QDs in solution phase.

## 3. Conclusion

COOH functionalized CdTe/PVA polymer matrix were prepared using simple and inexpensive method. These films were characterized through the XRD, UV–Visible and photoluminescence techniques. Appearance of strong absorption at 282 nm in functionalized CdTe/PVA thin film is observed for existing –COOH group of PVA. In photoluminescence, a strong and intense peak is observed at 520-540 nm in functionalized CdTe QDs in PVA polymer matrix. The close correlation in absorption data and luminescence data of CdTe QDs in liquid phase and CdTe in PVA polymer matrix validates the existence of nano sized QD structure in polymer matrix also. The smaller particle size in solid phase of the prepared functionalized CdTe in PVA polymer matrix (~ 6.2 nm from the absorption data and ~ 6 nm from the luminescence data) compared to that of the particle size in its solution phase (8 nm) validates the existence of agglomeration in liquid phase. The observed larger band gap of CdTe QD (9.5eV) in solid phase must be due to decrease in particle size in solid phase. As the CdTe QDs trapped in a solid polymer matrix the high intense and wide luminescence band shows a wide range of emmisivity in visible region. Due to increased band gap, CdTe QDs show much better electronic properties in solid phase than its liquid phase. So CdTe QDs show better optical and electrical responses in their solid phase. So CdTe QDs in solid form may be considered as the most promising candidates for applications as sensors for optoelectronic and biomedical industry.